\documentclass[aps,pre,twocolumn,groupedaddress]{revtex4-1}

\usepackage{graphicx}

\newcommand{\vind}[2]{v_{#1 #2,{\rm ind}}}
\newcommand{\veff}[2]{v_{#1 #2,{\rm eff}}}
\newcommand{\nden}[1]{n_{#1}}
\newcommand{\pd}[2]{\ensuremath\frac{\partial #1}{\partial #2}}

\def\Ns{N_s}
\def\nt{n_{\rm }}

\def\diel{\epsilon}

\def\rv{{\bf r}}                                                                               
                                                                               
\def\kv{{\bf k}}

\def\la{\left\langle}
\def\ra{\right\rangle}

\def\fp{F_p}
\def\fpm{F_m}

\def\kT{k_BT}
\def\lb{\lambda_B}
\def\iT{\beta}

\def\etal{{\it et al.}}

\begin{document}


\title{Effective electrostatic interactions in mixtures of charged colloids}


\author{Jun Kyung Chung and Alan R. Denton}
\email[]{alan.denton@ndsu.edu}
\affiliation{Department of Physics, North Dakota State University, Fargo, 
North Dakota 58108-6050, USA}


\date{\today}

\begin{abstract}
We present a theory of effective electrostatic interactions in polydisperse 
suspensions of charged macroions, generalizing to mixtures a theory previously
developed for monodisperse suspensions.  Combining linear response theory with
a random phase approximation for microion correlations, we coarse-grain the 
microion degrees of freedom to derive general expressions for effective 
macroion-macroion pair potentials and a one-body volume energy.  
For model mixtures of charged hard-sphere colloids, we give explicit analytical
expressions.  The resulting effective pair potentials have the same general form
as predicted by linearized Poisson-Boltzmann theory, but consistently incorporate
dependence on macroion density and excluded volume via the Debye screening constant.
The volume energy, which depends on the average macroion density, contributes to
the free energy and so can influence thermodynamic properties of deionized suspensions.
To validate the theory, we compute radial distribution functions of binary mixtures 
of oppositely charged colloidal macroions from molecular dynamics simulations of 
the coarse-grained model (with implicit microions), taking effective pair potentials 
as input.  Our results agree closely with corresponding results from more 
computationally intensive Monte Carlo simulations of the primitive model 
(with explicit microions).  Simulations of a mixture with large size and charge 
asymmetries indicate that charged nanoparticles can enhance electrostatic screening
of charged colloids.  The theory presented here lays a foundation for future large-scale
modeling of complex mixtures of charged colloids, nanoparticles, and polyelectrolytes.
\end{abstract}


\maketitle


\section{Introduction}\label{Introduction}

Soft materials, such as suspensions of colloids or nanoparticles and solutions
of polymers or surfactants, are complex mixtures of microscopic and mesoscopic 
components~\cite{Hamley2000}.  
Polydispersity in the intrinsic properties of macromolecules or mesoscopic particles
can significantly modify intermolecular (interparticle) forces~\cite{Israelachvili1992} 
and in turn self-assembly and macroscopic behavior.  While rigid particles have static
distributions of size and shape~\cite{Pusey1991,Evans1999}, polymer coils in solution 
can fluctuate in conformation~\cite{deGennes1979}.  Further variation can arise 
when counterions dissociate (in water or other polar solvents), leaving colloidal
or polyelectrolyte macroions with a broad charge distribution.  

The influence of polydispersity on thermodynamic phase behavior, structure, and
dynamics of soft materials has drawn increasing attention in recent years.
This trend stems not only from fundamental interest in the rich materials properties 
of mixtures, but also from the prevalence of polydispersity in natural colloids,
such as clays and many biological systems.  Moreover, tuning interparticle forces has 
practical applications in stabilizing unusual morphologies and engineering novel materials.  

Thermal and structural properties of bidisperse colloidal mixtures have been 
explored by a variety of experimental methods, including light scattering
and microscopy ~\cite{Klein1991-physica,Bartlett1990,Bartlett1992,Bartlett2000,Leunissen2005,
Hynninen2006,Royall2012,Klein1991-jpcm,Klein1992-jpcm}.  Theoretical and computational
studies have applied integral-equation methods~\cite{Klein1991-jpcm,Klein1992-jpcm,
Lowen1991-jpcm,Naegele1990,Klein1992-pra,Medina-Noyola2010,Medina-Noyola2011}, 
Poisson-Boltzmann theory~\cite{Torres2008-pre,Torres2008-jcp,castaneda-priego2011},
classical density-functional theory (DFT)~\cite{barrat1986,denton1990}, and 
computer simulations~\cite{Frenkel1991,Frenkel1993-1,Frenkel1993-2,Linse2005,
Dijkstra2007,Dijkstra2010}.
Recent related work has explored mixtures of colloids and nanoparticles, characterized
by extreme asymmetries in size and charge, via experiments~\cite{Lewis-2001-pnas,
Lewis-2001-langmuir,Lewis-2005-langmuir,Lewis-2008-langmuir}, 
integral-equation theory~\cite{Louis2004}, and simulation~\cite{Luijten2004}.

In modeling charged colloids, electrostatic interactions between macroions are 
commonly approximated by Yukawa (screened-Coulomb) effective pair potentials, 
as first derived in the classic works of Derjaguin and Landau~\cite{DL1941}
and Verwey and Overbeek~\cite{VO1948}, extending the Debye-H\"uckel theory of 
electrolytes.  Studies of charged colloidal mixtures also typically assume Yukawa 
pair potentials, which emerge from generalizing either the 
Derjaguin-Landau-Verwey-Overbeek (DLVO) theory or 
integral-equation theories based on the mean spherical approximation~\cite{Naegele1990}.
For salty suspensions, in which direct Coulomb interactions are strongly screened by 
microions (counterions and salt ions), the Yukawa model has proven reasonably accurate.
Recent observations of deionized mixtures~\cite{Royall2012}, however, have called into
question the accuracy of the Yukawa model when applied to weakly-screened macroions.

Previously, one of us modeled effective electrostatic interactions in one-component 
(monodisperse) suspensions of charge-stabilized colloids~\cite{Denton1999,Denton2000}
and polyelectrolyte solutions~\cite{Denton2003,Wang2004} using linear response theory. 
Within a mean-field (random-phase) approximation, equivalent to Poisson-Boltzmann theory
in its neglect of correlations between microions~\cite{DentonBook,Denton2010}, 
linear response theory recovers the usual Yukawa effective pair potential between
nonoverlapping macroions, but with a screening constant that depends on both 
salt and macroion densities and consistently incorporates excluded volume.
Beyond a density-dependent effective pair potential, the theory also predicts a 
one-body volume energy, as do related approaches to effective interactions~\cite{DentonBook}
based on integral-equation theories~\cite{Patey80,Belloni86,Khan87-mp,Khan87-pra,
Carbajal-Tinoco02,Petris02,Anta02,Anta03,Outhwaite02}, classical density-functional 
theory~\cite{vanRoij1997}, and extended Debye-H\"uckel theories~\cite{Chan85,Chan-pre01,
Chan-langmuir01,Warren2000}.  
Although independent of macroion coordinates, the volume energy contributes to the 
free energy a term that depends on macroion density and thus can affect bulk
thermodynamic properties at low salt concentrations (approaching counterion concentrations).  

The volume energy has been identified~\cite{vanRoij1997,Warren2000,Denton2006} as 
a possible origin of anomalous phase behavior observed in deionized monodisperse 
suspensions~\cite{Tata1992,Tata1997,Ise1994,Ise1996,Ise1999,Matsuoka1994,Matsuoka1996,
Matsuoka1999,Grier1997,Groehn2000,Royall2003}.  Theoretical modeling is complicated,
however, by nonlinear screening~\cite{Denton2004} and charge regulation~\cite{Denton2010,
castaneda-priego2011,castaneda-priego2007,Denton2008,Lu-Denton2010}.  
In a recent extension of the DFT approach, Bier \etal~\cite{Dijkstra2010} 
presented an expression for the volume energy of bidisperse charged colloids.  A subsequent 
experimental study~\cite{Royall2012} invoked this volume energy as a possible explanation of 
unusual fluid-crystal phase separation in deionized binary mixtures with large charge asymmetry.
Accurate theoretical predictions of the complex phase behavior of colloidal mixtures over a 
vast parameter space require a reliable theory of effective interactions.

In this paper we generalize linear response theory to polydisperse mixtures of macroions.
In Sec.~\ref{Model} we begin by defining the primitive model of charged colloids and
polyelectrolytes.  Within the primitive model, we develop in Sec.~\ref{Theory} the 
generalization of linear response theory to polydisperse mixtures and derive general 
expressions for the effective interactions.  In Sec.~\ref{AnalyticalResults} we present
explicit analytical expressions for the effective pair potentials and volume energies of
polydisperse suspensions of charged hard-sphere colloids and compare with previous 
theoretical results.  In Sec.~\ref{Structure} and the Appendix we discuss the 
calculation of structural and thermodynamic properties of bidisperse colloidal suspensions 
as functions of size and charge ratios.  Finally, in Sec.~\ref{Conclusions}
we summarize and conclude with suggestions for future applications.

\section{Primitive Model of Mixtures}\label{Model}
\begin{figure}[t]
\includegraphics{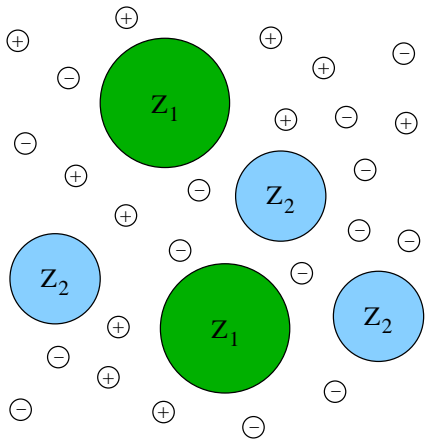}
\caption{Primitive model of binary mixture of charged colloids: two species of 
charged macroion (valences $Z_1$ and $Z_2$), microions (counterions and salt ions), 
and implicit solvent.
\label{modelCartoon}
}
\end{figure}

We consider spherical macroions of various species ($m=1,2,3,\ldots$), having diameters 
$\sigma_m$ (radii $a_m$) and valences $Z_m$, suspended in a solvent with microions
(species $\mu=1,2,3,\ldots$) of valences $z_{\mu}$ (see Fig.~\ref{modelCartoon}).  
Adopting the primitive model of charged colloids and polyelectrolytes, we treat the 
solvent as a dielectric continuum of dielectric constant $\epsilon$ that reduces 
the strength of electrostatic interactions.  The macroions are confined to a fixed 
volume $V$, while the microions (counterions, salt ions) are free to exchange with 
an electrolyte reservoir (e.g., via a semipermeable membrane), which maintains a 
fixed salt chemical potential (Donnan equilibrium) at absolute temperature $T$.  
For simplicity, we model the microions as point ions and assume a symmetric 
electrolyte of salt ion pairs with valences $z_+$ and $z_-$. 

The Hamiltonian of this model system can be separated according to 
$H = H_{\rm core} + H_{\rm el}$, where $H_{\rm core}$ incorporates interactions 
between macroion cores, as well as the total kinetic energy, and $H_{\rm el}$ 
is the total Coulomb electrostatic energy:
\begin{equation}
H_{\rm el} = H_m + H_{\mu} + H_{m\mu}~.
\end{equation} 
The first term on the right-hand side accounts for interactions among macroions ($m$), 
the second term interactions among microions ($\mu$), and the last term 
macroion-microion interactions.  An explicit expression for the macroion Hamiltonian is
\begin{equation}
H_m = \sum_m\sum_{i<j}^{N_m}v_{mm}(r_{ij})
+\sum_{m<n}\sum_{i=1}^{N_m}\sum_{j=1}^{N_n}v_{mn}(r_{ij})~,
\label{Hm}
\end{equation}
where $N_m$ is the number of macroions of species $m$ and 
$v_{mn}(r_{ij})=Z_mZ_ne^2/\epsilon r_{ij}$ is the (Coulomb) potential energy 
between a pair of macroions (labeled $i$ and $j$) of species $m$ and $n$
separated by center-to-center distance $r_{ij}$, $e$ being the electron charge.
Similarly, the microion Hamiltonian is
\begin{equation}
H_{\mu} = \sum_{\mu}\sum_{i<j}^{N_{\mu}}v_{\mu\mu}(r_{ij})
+\sum_{\mu<\nu}\sum_{i=1}^{N_{\mu}}\sum_{j=1}^{N_{\nu}}v_{\mu\nu}(r_{ij})~,
\label{Hmu}
\end{equation}
where $N_{\mu}$ is the number of microions of species ${\mu}$ and 
$v_{\mu\nu}(r_{ij})=z_{\mu}z_{\nu}e^2/\epsilon r_{ij}$ is the potential energy 
between a pair of microions of species ${\mu}$ and ${\nu}$.
Finally, the macroion-microion interaction Hamiltonian is given by
\begin{equation}
H_{m\mu} = \sum_{m,\mu}\sum_{i=1}^{N_m}\sum_{j=1}^{N_{\mu}}v_{m\mu}(r_{ij})~,
\label{Hmmu}
\end{equation}
where $v_{m\mu}(r_{ij})=Z_mz_{\mu}e^2/\epsilon r_{ij}$ is the macroion-microion
pair potential energy.
Latin and Greek subscripts here refer to macroions and microions, respectively.
Note that the subscripts $m$ and $\mu$ are used both to represent macroions and microions 
as a whole and as an index to label different species of macroion and microion, 
the distinction being clear from the context.  The condition of global electroneutrality 
dictates that $\sum_m Z_m N_m+\sum_{\mu}z_{\mu}N_{\mu}=0$.

\section{Linear Response Theory}\label{Theory}
Within the primitive model, we outline a general coarse-graining approach to 
modeling effective electrostatic interactions in polydisperse suspensions of 
charged macroions, extending to mixtures the linear response theory formulated 
previously for monodisperse suspensions of spherical macroions~\cite{Denton1999,
Denton2000}.  
Integrating out microion degrees of freedom from the partition function, assuming
linear response of microion densities to macroion electrostatic potentials, and 
making a mean field approximation for the microion response functions, we obtain 
effective pair potential energies between macroion species $m$ and $n$ of the form
\begin{equation}
\veff{m}{n}(r)=A_{mn}\frac{\exp(-\kappa r)}{r}~, \quad r > a_m + a_n~,
\label{crossnp}
\end{equation}
where $\kappa$ is the inverse Debye screening length and the prefactors $A_{mn}$ 
depend on macroion sizes and charges.  Equation~(\ref{crossnp}) is the well-known Yukawa 
effective pair potential assumed in many simulation studies~\cite{Allahyarov1998,
Allahyarov2009,Hynninen2006}.
In addition to confirming the general form of the effective pair potentials, however,
our approach also incorporates macroion excluded volume into the screening constant
and yields a one-body volume energy, which depends on the bulk densities of 
{\it all} microions (both salt ions and counterions).

\subsection{Coarse graining microion degrees of freedom}
We begin by extending to mixtures a general statistical mechanical procedure  
for formally averaging over microion degrees of freedom such that the system 
partition function remains unchanged.  If this averaging is performed exactly,
the resulting coarse-grained model will reproduce thermodynamic properties 
of the original model \cite{DentonBook,Likos2001}. 
The canonical partition function for our model mixture is given by
\begin{equation}
{\cal Z} = \la\la\exp(-\iT H)\ra_{\mu}\ra_m~,
\label{fulltrace}
\end{equation}  
where $\iT\equiv 1/k_BT$ and the angular brackets represent classical traces 
over relevant degrees of freedom.  After coarse graining, achieved by integrating 
over microion degrees of freedom for a fixed macroion configuration, 
Eq.~(\ref{fulltrace}) can be re-expressed as
\begin{equation}
{\cal Z} = \la \exp(-\iT H_{\rm eff})\ra_m~,
\label{partialTr}
\end{equation}  
where $H_{\rm eff} \equiv H_{\rm{core}} + H_m + F_{\mu}$ and
\begin{equation}
F_{\mu} \equiv - \kT \ln\la\exp\left[-\iT (H_{\mu} + H_{m\mu})\right]\ra_{\mu}
\end{equation}
can be interpreted as the free energy of microions in a fixed configuration 
of macroions. To make coarse-grained models of effective interactions practical
for simulations or further theoretical development, approximations are necessary 
to render $F_{\mu}$ in an analytical or numerically computable form.

\subsection{Linear response approximation for microions 
\label{ResponseTheory}}
Following the general approach of Silbert and co-workers~\cite{Silbert1,Silbert2,Silbert3}, 
we regard the interactions of the macroions with the microions as external perturbations 
to a uniform microion plasma.  As a first step, we define an intermediate free energy 
as a function of a charging parameter $\lambda$,
\begin{equation}
F_{\mu}(\lambda) \equiv - \kT \ln\la\exp\left[-\iT (H_{\mu} +
                          \lambda H_{m\mu})\right]\ra_{\mu}~.
\end{equation}
With this definition, $F_{\mu}=F_{\mu}(\lambda=1)$ can be written as
\begin{equation}
F_{\mu} = F_{\mu}(0) + 
          \int_{0}^{1} d\lambda\, \la H_{m\mu} \ra_{\lambda}~,
\label{perturbF}
\end{equation}
where $\la~\ra_{\lambda}$ denotes an average over microion degrees of freedom in 
a system where the macroions are charged to a fraction $\lambda$ of their full charges.  

In Eq.~(\ref{perturbF}), $F_{\mu}(0)$ is the free energy of a reference system 
consisting of a classical gas of microions in a free volume $V_f=V(1-\eta)$, 
which excludes the fraction 
\begin{equation}
\eta = \frac{4\pi}{3V}\sum_m N_m a_m^3 
\label{eta}
\end{equation}
of the total volume that is occupied by macroion hard cores. To ensure that the
reference system is electroneutral, it is convenient to add to and subtract from
$F_{\mu}(0)$ the energy of a uniform compensating background charge distribution,
occupying the same free volume, having uniform number density 
\begin{equation}
\rho_b = \frac{1}{V_f}\sum_m Z_mN_m~.
\end{equation}
Denoting the energy of this background by 
\begin{equation}
E_b = -\frac{V_f\rho_b^2}{2\diel} \lim_{k\to 0} \frac{4\pi e^2}{k^2}~,
\end{equation}
we can redefine the microion interaction energies as
\begin{equation}
H_{\mu}' \equiv H_{\mu} + E_b~, \qquad
H_{m \mu }' \equiv H_{m \mu } - E_b~.
\end{equation}
The microion free energy $F_{\mu}$ then can be expressed as
\begin{equation}
F_{\mu} = \fp + \int_{0}^{1} d\lambda\, \la H_{m\mu}' \ra_{\lambda}~,
\label{perturbF2}
\end{equation}
where $\fp = -k_BT\ln\la\exp(-\iT H_{\mu}')\ra_{\mu}$ is the free energy 
of a microion plasma with the neutralizing background charge density $e\rho_b$.

The next step in approximating $F_{\mu}$ is to relate the macroion-microion
Hamiltonian [Eq.~(\ref{Hmmu})] to number density operators $\rho_m({\rv})$ 
and $\rho_{\mu}({\rv})$ of macroions and microions, respectively, and to 
the macroion-microion pair potentials $v_{m\mu}(r)$: 
\begin{equation}
H_{m\mu} = \sum_m\sum_{\mu}\int_{V_f}d{\rv} \int_{V_f}d{\rv}'\,
\rho_m({\rv}) v_{m\mu}(|{\rv}-{\rv}'|) \rho_{\mu}({\rv}')~.
\end{equation}
The integrand in Eq.~(\ref{perturbF2}) then can be expressed in terms of 
Fourier components:
\begin{equation}
\la H_{m\mu}' \ra_{\lambda} = \frac{1}{V_f} 
  \sum_{m} \sum_{\mu} \sum_{\kv} 
\hat \rho_{m}(\kv) \hat v_{m\mu}(k)
\la \hat \rho_{\mu}(-\kv) \ra_{\lambda} - E_b~,
\label{perturbH}
\end{equation}
where the Fourier transforms are defined over the free volume, for example,
\begin{equation}
\hat \rho_{m}(\kv) = \int_{V_f} d\rv\, \rho_{m}(\rv)\exp(-i\kv\cdot\rv)~.
\end{equation}

To develop a response theory, we first define an external potential applied by 
the macroions to the (otherwise uniform) microion plasma:   
\begin{equation}
v_{{\rm ext}}(\rv) \equiv
\sum_{m}^{} Z_{m} \int d\rv'\, v_{m}(|\rv-\rv'|)
                      \rho_{m}(\rv')~,
\end{equation}
where $v_m(r) \equiv v_{m\mu}(r)/Z_m z_{\mu}$.  We then make the approximation 
that the microion densities respond {\it linearly} to the macroion external potential.
Denoting by $\chi_{\mu\nu}(k)$ the linear response functions of the unperturbed 
microion plasma (with $\lambda = 0$), and defining
$\chi_{\mu}(k) \equiv \sum_{\nu} z_{\nu}\chi_{\mu\nu}(k)$, 
then to {\it linear} order in the external potential
\begin{equation}
\la \hat \rho_{\mu}(\kv) \ra_{\lambda} = \lambda \ \chi_{\mu}(k)
\hat v_{{\rm ext}}(\kv), \qquad k\neq 0~,
\label{lrapprox}
\end{equation}
the Fourier transform of the external potential being
\begin{equation}
\hat v_{{\rm ext}}(\kv) = \sum_{m}^{} Z_{m} 
\hat v_{m}(k) \hat \rho_{m}(\kv)~.
\label{vkext}
\end{equation}
Note that the $k=0$ component must be excluded since $\hat\rho_{\mu}(0)=N_{\mu}$ 
is fixed by the condition of electroneutrality.

Using Eqs.~(\ref{perturbH}) and (\ref{lrapprox}), the linear response approximation
for the microion free energy [Eq.~(\ref{perturbF2})] can be expressed as
\begin{eqnarray}
F_{\mu} &=& \fp + \frac{1}{2V_f}\sum_m \sum_{\mu} 
\sum_{\kv \neq 0} \hat \rho_m(\kv) \hat v_{m\mu}(k)
\chi_{\mu}(k) \hat v_{{\rm ext}}(-\kv) \nonumber \\
&+& \frac{1}{V_f}\sum_{m} \sum_{\mu} 
N_mN_{\mu} \lim_{k\to 0} \hat v_{m\mu}(k) - E_b~.
\label{lrFree}
\end{eqnarray}
Equation~(\ref{lrFree}) can be recast in the more intuitive form 
\begin{equation}
F_{\mu} =
\sum_m\sum_{i<j}^{N_m}\vind{m}{m}(r_{ij})
+ \sum_{m<n}\sum_{i=1}^{N_m}\sum_{j=1}^{N_n}\vind{m}{n}(r_{ij})
+ E_0~,
\label{lrFree2}
\end{equation}
where $\vind{m}{n}(r)$ are microion-induced pair potentials between macroions, 
whose Fourier transforms are given by
\begin{equation}
\hat \vind{m}{n}(k) = Z_mZ_n \hat v_m(k) \hat v_n(k)
\sum_{\mu}z_{\mu}\chi_{\mu}(-k)~,
\label{fourier_np}
\end{equation}
and $E_0$ is a one-body volume energy:
\begin{eqnarray}
E_0 &=& \fp +\frac{1}{2}\sum_{m} N_m \lim_{r\to 0}\vind{m}{m}(r) 
\nonumber \\ 
&-&\frac{1}{2V_f}\sum_{m,n} N_mN_n \lim_{k\to 0} \hat \vind{m}{n}(k) 
\nonumber \\ 
&+& \frac{1}{V_f}\sum_{m} \sum_{\mu} N_{m}N_{\mu}\lim_{k\to 0} \hat v_{m\mu}(k) - E_b~.
\label{volE}
\end{eqnarray}
Equation~(\ref{lrFree2}) suggests expressing the effective Hamiltonian as
\begin{eqnarray}
H_{\rm eff} &=& H_{\rm{core}} + \sum_m\sum_{i<j}^{N_m} \veff{m}{n}(r_{ij})
\nonumber \\ 
&+& \sum_{m<n}\sum_{i=1}^{N_m}\sum_{j=1}^{N_n} \veff{m}{n}(r_{ij}) + E_0~,
\label{effH2}
\end{eqnarray}
thus identifying
\begin{equation}
\veff{m}{n}(r) = v_{mn}(r)+\vind{m}{n}(r)
\label{vmneff}
\end{equation}
as an effective (microion-mediated) pair potential between macroions
of species $m$ and $n$.  

Note that our coarse-grained model involves only one- and two-body effective
interactions, which is a direct consequence of the linear approximation for 
the response of the microion densities [Eq.~(\ref{lrapprox})].  Nonlinear 
response entails many-body effective interactions, as well as corrections 
to the one- and two-body interactions~\cite{Denton2004}.
The linear response approximation is reasonable for sufficiently weakly charged
macroions and proves valid even for highly charged macroions if the bare valence
is replaced by an {\it effective} valence via charge renormalization 
theory~\cite{Denton2008,Denton2010,Lu-Denton2010}.  For monodisperse suspensions,
the theory accurately predicts thermodynamic and structural properties 
(osmotic pressures and radial distribution functions) for electrostatic coupling 
strengths as high as $Z_m\lambda_B/a_m\simeq 15$~\cite{Denton2008,Denton2010,Lu-Denton2010}.

\section{Analytical Results}\label{AnalyticalResults}
Calculating effective interactions in polydisperse mixtures of charged colloids
requires approximating the linear response functions $\chi_{\mu}(k)$.  Following
previous studies of monodisperse charged colloids~\cite{Denton1999,Denton2000}, 
we adopt the random-phase approximation, which provides $\chi_{\mu}(k)$ in
analytical form and thus yields analytical expressions for the induced pair potentials
between macroions, from Eq.~(\ref{fourier_np}), and for the volume energy, 
from Eq.~(\ref{volE}).

\subsection{Response functions of the microion plasma}
The linear response functions of the reference microion plasma 
are proportional to the corresponding partial structure factors 
\cite{HansenMcDonald}: 
\begin{equation}
\chi_{\mu\nu}(k) = -\iT \sum_{\mu} n_{\mu}S_{\mu\nu}(k)~,
\label{chimunu}
\end{equation}
where $\nden{\mu}=N_{\mu}/V_f$ is the average number density of microion species
$\mu$ in the {\it free volume}, thus incorporating theexcluded volume of macroion 
hard cores.  The partial structure factors $S_{\mu\nu}(k)$ are related in turn 
to the Fourier transforms of the pair correlation functions $h_{\mu\nu}(r)$:
\begin{equation}
S_{\mu\nu}(k) = x_{\mu}\left[\delta_{\mu\nu} + \nden{\nu} \hat h_{\mu\nu}(k)\right]~,
\label{partialSk}
\end{equation}
where $x_{\mu}$ is the concentration of microion species $\mu$.  In Fourier space,
$\hat h_{\mu\nu}(k)$ is related to the direct correlation function 
$\hat c_{\mu\nu}(k)$ via the Ornstein-Zernike integral equation 
\begin{equation}
\hat h_{\mu\nu}(k) = \hat c_{\mu\nu}(k) + \sum_{\lambda} \nden{\lambda} 
\hat c_{\mu\lambda}(k) \hat h_{\lambda\nu}(k)~.
\end{equation}
For a weakly coupled plasma, whose average Coulomb energy is much lower than 
the average thermal energy, we can approximate the 
direct correlation functions by their asymptotic limits  
$\hat c_{\mu\nu}(k) \simeq -\iT \hat v_{\mu\nu}(k)=z_{\mu}z_{\nu}\hat c(k)$, 
where $\hat c(k)={-4\pi\lb}/{k^2}$ and $\lb=e^2/\diel k_BT$ is the Bjerrum length,
defined as the separation between two elementary charges $e$ at which the 
electrostatic potential energy equals the typical thermal energy $k_BT$. 
Further assuming 
$\hat h_{\mu\nu}(k)=z_{\mu}z_{\nu} \hat h(k)$, it follows that 
\begin{equation}
\hat h_{\mu\nu}(k)=\frac{z_{\mu}z_{\nu} \hat c(k)}{1-n_0\hat c(k)}
\label{dcMSA}
\end{equation}
with $n_0\equiv\sum_{\mu}z_{\mu}^2\nden{\mu}$.
Combining Eqs.~(\ref{chimunu})-(\ref{dcMSA}), we obtain the linear response
functions 
\begin{equation}
\chi_{\mu}(k) = -\frac{\beta z_{\mu}\nden{\mu}}{1+\kappa^2/k^2}~,
\label{chiMSA}
\end{equation}
where the inverse Debye screening length is defined as
$\kappa\equiv\sqrt{4\pi\lb n_0}$.
We emphasize that $\kappa$ here incorporates the macroion excluded volume, 
since $n_0$ involves the microion densities $n_{\mu}$ in the free volume,
i.e., the volume not excluded by the macroion hard cores.
Thus, our definition of $\kappa$ is larger than the conventional definition 
by a factor of $1/\sqrt{1-\eta}$.
With Eq.~(\ref{chiMSA}), the effective electrostatic interactions now can be
explicitly calculated.

\subsection{Effective pair potentials and volume energy}
The general expressions derived for the effective pair potentials and volume energy 
apply to any type of spherical macroion, provided only that the macroion-microion
interaction can be factorized as $v_{m\mu}(r)=Z_m z_{\mu} v_m(r)$.  
For separations exceeding the macroion radius (assuming point microions),
$v_{m\mu}(r)$ is of Coulomb form.  For colloidal macroions with an impenetrable 
core, the potential inside the core may be {\it chosen}~\cite{vanRoij1997,Denton1999} 
to ensure exclusion of microions from the core:
\begin{equation}
\beta v_m(r)~=~\lb\left\{ \begin{array}
{l@{\quad\quad}l}
\frac{\displaystyle 1}{\displaystyle r}, & r > a_m \\[1ex]
\frac{\displaystyle \alpha_m}{\displaystyle a_m}~, 
& r < a_m~, \end{array} \right.
\label{vmacmic}
\end{equation}
where the constant $\alpha_m$ can be fixed to impose the condition
$\rho_{\mu}(\rv)=0$ for $r<a_m$.  With the appropriate choice of 
$\alpha_m = \kappa a_m/(1+\kappa a_m)$, Eq.~(\ref{vmacmic}) has the Fourier transform 
\begin{equation}
\beta\hat v_m(k) = \frac{4\pi\lb}{k^2}\frac{1}{1+\kappa a_m}
\left[\cos(ka_m)+\kappa\frac{\sin(ka_m)}{k}\right]~.
\label{hatvm}
\end{equation}
Next, substitution of Eqs.~(\ref{chiMSA}) and (\ref{hatvm}) into Eq.~(\ref{fourier_np})
yields the Fourier transform of the microion-induced pair potential
\begin{equation}
\beta\hat \vind{m}{n}(k) = -\frac{Z_mZ_n\beta^2\kappa^2}{4\pi\lb}
~\frac{k^2}{k^2+\kappa^2}~\hat v_m(k)\hat v_n(k)~,
\label{FTvindmn}
\end{equation}
with an inverse transform 
\begin{widetext}
\begin{eqnarray}
\beta\vind{m}{n}(r) =
\left\{ \begin{array}
{l@{\quad\quad}l}
B_{mn}\frac{\displaystyle \exp[-\kappa(r-a_m-a_n)]}{\displaystyle \kappa r}
-\beta v_{mn}(r)~,
\qquad \qquad \quad r \ge a_m+a_n \\[2ex]
B_{mn}\left\{ \begin{array}
{l@{\quad\quad}l}
{\displaystyle -\frac{\kappa}{2}(a_m+a_n-\vert a_m-a_n \vert)-1}~, 
& r \le \vert a_m-a_n\vert  
\\[2ex]
{\displaystyle 
\frac{\kappa}{4} \left[r+\frac{(a_m-a_n)^2}{r}-2(a_m+a_n)\right]-1}~,
& \vert a_m-a_n\vert < r < a_m+a_n~, 
\end{array} \right.
\end{array} \right.
\hspace*{-3.5cm}
\label{vindmnr}
\hspace*{2cm} 
\end{eqnarray}
\end{widetext}
where $B_{mn}\equiv Z_mZ_n\kappa\lb/[(1+\kappa a_m)(1+\kappa a_n)]$.
Substituting this result for the induced pair potentials into Eq.~(\ref{vmneff}),
we finally obtain effective macroion-macroion pair potentials (for $r\ge a_m+a_n$)
\begin{equation}
\beta\veff{m}{n}(r) = Z_mZ_n\lb
~\frac{\exp[\kappa(a_m+a_n)]}{(1+\kappa a_m)(1+\kappa a_n)}
~\frac{\exp(-\kappa r)}{r}~. 
\label{veffmn}
\end{equation}
Thus, we recover the Yukawa pair potential of Eq.~(\ref{crossnp}), 
with the prefactor determined to be
\begin{equation}
A_{mn} = Z_mZ_n\frac{e^2}{\epsilon}~
\frac{\exp[\kappa(a_m+a_n)]}{(1+\kappa a_m)(1+\kappa a_n)}~.
\label{Amn}
\end{equation}
The effective pair potentials of Eq.~(\ref{veffmn}) are the same as those predicted 
by the DLVO theory extended to mixtures in the dilute limit, i.e., by solving the 
linearized Poisson-Boltzmann equation with free boundary conditions.  Our result 
applies also, however, at nonzero macroion concentrations --- as long as the 
linear response approximation remains valid --- in which case the screening constant
depends on both salt and macroion densities and incorporates the macroion excluded volume.

Similar results for effective pair potentials in colloidal mixtures have been 
derived by Ruiz-Estrada~\etal~\cite{Naegele1990} using integral-equation theory.  
Starting from the primitive model, and contracting the Ornstein-Zernike equation
(relating pair and direct correlation functions) to eliminate explicit reference
to the direct correlation functions between microions, these authors obtain a 
formal expression for effective direct correlation functions between macroions.
Making a mean spherical approximation (MSA) for all correlation functions, they 
obtain an analytical expression of the same general Yukawa form as Eq.~(\ref{crossnp}).  
The effective pair potentials derived from the MSA [Eqs.~(2.15) and (2.16) in 
Ref.~\cite{Naegele1990}] differ, however, from ours [Eq.~(\ref{veffmn})] in two respects.
First, the prefactors are different, the MSA result reducing to our $A_{mn}$ 
only in the dilute limit.  Second, the MSA expression for the screening constant 
[Eq.~(2.7) in Ref.~\cite{Naegele1990}], like that in the DLVO theory, 
does not incorporate the macroion excluded volume.

Beyond effective pair potentials, the linear response approach also consistently
yields a one-body volume energy.  By substituting Eqs.~(\ref{hatvm})-(\ref{vindmnr}) 
into Eq.~(\ref{volE}), we arrive at an explicit result for the volume energy of 
a colloidal mixture:
\begin{equation}
\iT E_0~=~\iT \fp -
\frac{\lb}{2} \sum_{m} \frac{N_m Z_m^2}{a_m+\kappa^{-1}} -
\frac{1}{2}
\frac{\left( \sum_{m}Z_m N_m \right)^2}{\sum_{\mu}z_{\mu}^2N_{\mu}}~.
\label{volEexplicit}
\end{equation}
Assuming a weakly coupled microion plasma, the first term on the right-hand side 
can be approximated as the free energy of an ideal gas of microions:
\begin{equation}
\iT \fp \simeq \sum_{\mu}^{} N_{\mu}[\ln(\nden{\mu}\Lambda_{\mu}^3)-1]~,
\label{micFree}
\end{equation}
$\Lambda_{\mu}$ being the thermal wavelength of microion species $\mu$.
The second term on the right-hand side of Eq.~(\ref{volEexplicit}) represents
the self energy of the macroions embedded in the microion plasma.
A similar expression for the volume energy of colloidal mixtures can been
derived from the DFT approach to effective interactions~\cite{Dijkstra2010}.  
Our result for $E_0$ differs, however, in the manner in which 
macroion excluded volume is incorporated via the screening constant.

\section{Structure and Thermodynamics}\label{Structure}

\begin{figure}[t]
\includegraphics[width=0.45\textwidth]{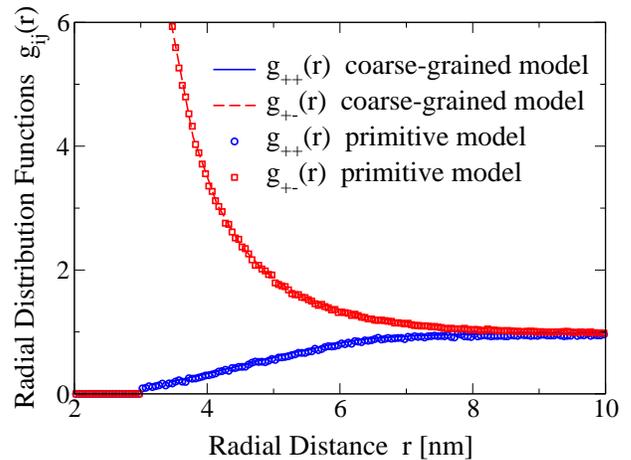}
\caption {Radial distribution functions from molecular dynamics simulations 
of the coarse-grained model (curves) compared with corresponding results from
Monte Carlo simulations~\cite{Linse2005} of the primitive model (symbols) 
for a salt-free binary mixture of oppositely charged ($Z_+=-Z_-=5$), equally sized 
($\sigma_+=\sigma_-=3$ nm) colloids at equal volume fractions ($\eta_+=\eta_-=0.005319$).
\label{oppChargedColloids}}
\end{figure}

\subsection{Pair structure of binary mixtures}\label{structure}
To validate the linear response theory and assess the accuracy of the predicted 
effective pair potentials, we performed molecular dynamics (MD) simulations of 
the coarse-grained model (with implicit microions).  Using the LAMMPS package 
\cite{Plimpton1995}, we computed macroion-macroion radial distribution functions 
(RDFs) $g_{ij}(r)$ and compared with available results from Monte Carlo (MC) 
simulations \cite{Linse2005} of a binary mixture of {\it oppositely} charged,
equally sized macroions in the primitive model (with explicit counterions) in
a salt-free aqueous suspension.  
For a direct comparison, we chose the same system parameters as in Ref.~\cite{Linse2005}: 
hard-sphere diameters $\sigma_+=\sigma_-=3$ nm, valences $Z_+=-Z_-=5$, and 
volume fractions $\eta_+=\eta_-=0.005319$.

For convenience, in our MD simulations, we replaced the hard-sphere interactions
between macroions with the repulsive part of the Lennard-Jones pair potential, 
$v_{\rm LJ}(r)=4\epsilon_{\rm LJ}\left[(\sigma_{\rm LJ}/r)^{12} -(\sigma_{\rm LJ}/r)^6\right]$, 
cut and shifted to zero at its minimum, which we matched to the diameter of the 
colloids: $\sigma_c=2^{1/6}\sigma_{\rm LJ}$.  We set $\epsilon_{\rm LJ}=5000$ kcal/mol,
checking that higher values did not significantly affect the RDFs, and cut and
shifted to zero the effective pair potentials [Eq. (\ref{veffmn})] at
$r_{\rm cut}=20/\kappa$, beyond which range the interactions are negligible.

Starting from initial configurations of 4000 particles on a face-centered cubic 
lattice, with appropriate concentrations of each species, we performed simulations 
in the canonical ensemble at fixed temperature ($T=298$ K) with periodic boundary 
conditions in a cubic simulation box of side length $L$ chosen to ensure that 
$L/2 > r_{\rm cut}$.  Following an initial equilibration phase, we sampled 
configurations and collected statistics at regular intervals over $10^6$ time steps.

As seen in Fig.~\ref{oppChargedColloids}, the macroion-macroion RDFs calculated 
for this system from our simulations of the coarse-grained model are in excellent 
agreement with those obtained from MC simulations of the primitive model.  We 
caution, however, that the electrostatic coupling in this system, characterized by 
$Z\lambda_B/\sigma=1.2$, is relatively weak.  Preliminary comparisons indicate 
that more strongly coupled systems ($Z\lambda_B/\sigma>3$) must be modeled using 
effective macroion charges consistently derived from charge renormalization 
theory~\cite{ChungDentonUnpublished}.

\begin{figure}[h]
\vspace*{0.4cm}
\includegraphics[width=0.45\textwidth]{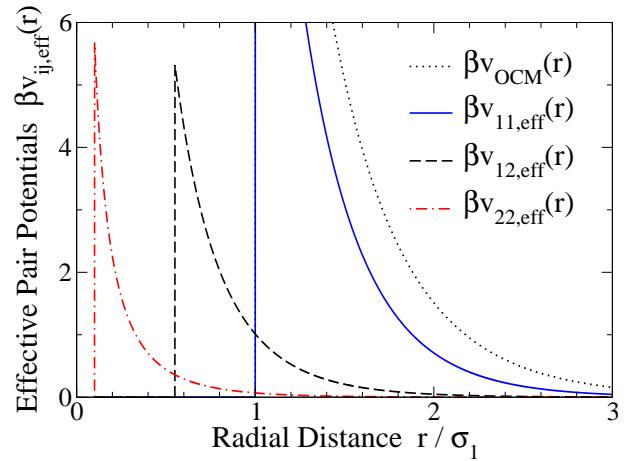}
\caption {Effective pair potentials of a salt-free aqueous suspension of macroions with
radii $a_1=50$ nm and $a_2=5$ nm, valences $Z_1=100$ and $Z_2=10$, concentration $N_1/N_2=1/3$,
and volume fractions $\eta_1=0.2$ and $\eta_2=0.0006$ [from Eqs.~(\ref{veffmn}) and 
(\ref{Amn})].  Curves represent (left to right) $\beta v_{22,{\rm eff}}(r)$ (dot-dashed)
$\beta v_{12,{\rm eff}}(r)$ (dashed), and $\beta v_{11,{\rm eff}}(r)$ (solid).
The dotted curve is the effective pair potential of the one-component model (OCM) of 
the same suspension in the absence of the smaller macroions (species 2).
\label{pairPotentials}}
\end{figure}

\begin{figure}[h]
\includegraphics[width=0.45\textwidth]{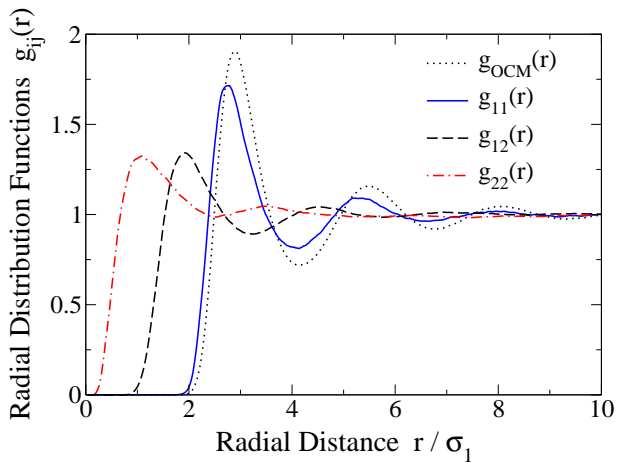}
\caption {Radial distribution functions from molecular dynamics simulations of 
the coarse-grained model of a salt-free aqueous suspension of macroions with
radii $a_1=50$ nm and $a_2=5$ nm, valences $Z_1=100$ and $Z_2=10$, concentration $N_1/N_2=1/3$,
and volume fractions $\eta_1=0.2$ and $\eta_2=0.0006$.  
Curves represent (main peaks, left to right) $g_{22}(r)$ (dot-dashed)
$g_{12}(r)$ (dashed), and $g_{11}(r)$ (solid).
The dotted curve is the RDF of the one-component model of the same suspension
in the absence of the smaller macroions.
\label{likeChargedColloids}}
\end{figure}

To demonstrate an application to a mixture that is bidisperse in both size and charge,
and to explore the influence of nanoparticles on the structure of colloids, 
we performed an MD simulation of a mixture with relatively large size and charge 
asymmetries.  Specifically, we simulated the coarse-grained model of a salt-free 
aqueous suspension of $N_1=500$ colloids, of radius $a_1=50$ nm and valence $Z_1=100$,
and $N_2=1500$ nanoparticles, of radius $a_2=5$ nm and valence $Z_2=10$, 
at volume fractions $\eta_1=0.2$ and $\eta_2=0.0006$.  Figures~\ref{pairPotentials} 
and \ref{likeChargedColloids} show, respectively, the effective pair potentials 
[from Eqs.~(\ref{veffmn}) and (\ref{Amn})] and the corresponding RDFs from our
simulations of this model colloid-nanoparticle mixture.  For comparison, results are
shown both for the mixture and for a one-component suspension of type-1 macroions only.
Evidently, the smaller (nano) particles act to soften the pair interactions, and 
correspondingly weaken pair correlations, between the larger particles.  We interpret
the role of the nanoparticles as enhancing screening of the charged colloids.

To assess the significance of the excluded-volume correction to the inverse 
Debye screening constant $\kappa$, and hence to the effective pair potentials, 
we performed a test simulation using uncorrected pair potentials for the same
colloid-nanoparticle mixture.  Even for such a concentrated suspension,
the excluded-volume correction only slightly reduces the amplitude and range 
of the effective pair potentials.  The resulting RDFs are, consequently, 
barely distinguishable from those shown in Fig.~\ref{likeChargedColloids}.  
The excluded-volume correction thus has a relatively minor impact on macroion
pair structure.  However, the same correction alters the density dependence 
of the effective interactions --- both the effective pair potentials {\it and}
the one-body volume energy --- which can significantly modify bulk thermodynamic
properties, such as osmotic pressure, as shown in Sec.~\ref{pressure}.

\subsection{Pressure and equation of state}\label{pressure}
The pressure of a colloidal mixture can be computed from the Helmholtz free energy $F$
via $p=-(\partial F/\partial V)_{N_m,N_s}$, where the subscripts denote fixing of 
all macroion and salt ion numbers (fixed $T$ is implied).  Equivalently, 
$p=n^2(\partial (F/N)/\partial n)_{x_m,x_s}$, where $N$ and $n=N/V$ are the 
total macroion number and number density, $x_m=N_m/N$ is the concentration 
of macroion species $m$, and $x_s=N_s/N$ is the salt concentration.

The Helmholtz free energy of the system naturally divides into two parts,
$F = E_0 + \fpm$, where $E_0$ is the volume energy arising from tracing out 
the microion degrees of freedom and $\fpm$ is the free energy associated with 
effective interactions between macroions.  Correspondingly, the pressure 
can be separated as $p=p_0+p_m$, where the volume energy contribution
[Eq.~(\ref{volEexplicit})] is given by
\begin{eqnarray}
\iT p_0 &=& \nt^2\iT \left(\pd{(E_0/N)}{\nt}\right)_{x_m,x_s}
\nonumber\\
&=& \sum_{\mu}n_{\mu} - \frac{\kappa \lb}{4(1-\eta)}\sum_{m}^{}
\frac{n_m Z_m^2}{(1+\kappa a_m)^2}
\label{pvolE}
\end{eqnarray}
and the macroion contribution is given by
\begin{equation}
\iT p_m = \sum_{m}^{}\nden{m} -
\iT\la\left(\pd{U}{V}\right)_{x_m,x_s} \ra~.
\label{pMac}
\end{equation}
Here $n_m=N_m/V$ denotes the number density of macroion species $m$ and 
\begin{equation}
U = \sum_m\sum_{i<j}^{N_m}\veff{m}{m}(r_{ij})
+ \sum_{m<n}\sum_{i=1}^{N_m}\sum_{j=1}^{N_n}\veff{m}{n}(r_{ij})
\label{U}
\end{equation}
is the potential energy associated with macroion pair interactions.
The ensemble average of $\partial U/\partial V$ can be approximated by 
either a perturbation theory or molecular simulations, taking into account the 
dependence of the effective pair potentials on the macroion and salt densities 
\cite{Louis2002,castaneda-priego2006,Lu-Denton2007}.  As shown in the Appendix,
this density dependence results in extra terms in addition to the usual virial term.
Taken together, Eqs.~(\ref{pvolE}) and (\ref{pMac}) can be used to calculate
the pressure of a polydisperse colloidal suspension or polyelectrolyte solution.

\begin{figure}[h]
\vspace*{0.5cm}
\includegraphics[width=0.45\textwidth]{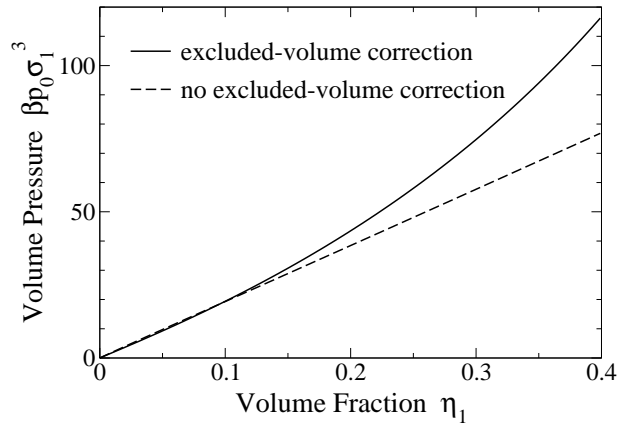}
\caption {Contribution to the osmotic pressure from the one-body volume energy 
[calculated from Eq.~(\ref{pvolE})] for the suspension whose parameters are specified
in the caption to Fig.~\ref{likeChargedColloids}.  Solid and dashed curves represent, 
respectively, predictions with and without excluded volume accounted for in the 
volume energy.
\label{pvol}}
\end{figure}

Finally, to illustrate the significance for thermodynamic properties of the
excluded-volume correction to the effective interactions, we examine the volume energy
contribution $p_0$ [Eq.~(\ref{pvolE})] to the total osmotic pressure of the 
colloid-nanoparticle mixture described in Sec.~\ref{structure} (see caption to 
Fig.~\ref{likeChargedColloids}).
Figure~\ref{pvol} shows the concentration dependence of $p_0$, both with and without
excluded volume taken into account.  Evidently, with increasing macroion concentration,
the excluded-volume correction increasingly affects the osmotic pressure, which in turn
can influence thermodynamic phase behavior.

\section{Conclusions}\label{Conclusions}
In summary, we have presented a theory of effective electrostatic interactions 
for polydisperse suspensions of charged macroions, thus generalizing to mixtures 
a theory previously developed for monodisperse suspensions.  Within a coarse-graining 
framework that integrates out microion degrees of freedom, we derived general expressions 
for effective macroion-macroion pair potentials and a one-body volume energy.  
The theory is based on a linear response approximation for the microion densities
and a mean-field random phase approximation for microion structure that neglects 
all but long-range microion correlations.  For model mixtures of charged hard-sphere 
colloids, we have presented explicit analytical expressions for the effective interactions.  
These expressions should be accurate for suspensions of weakly correlated (monovalent) 
microions and macroions whose charges are sufficiently low that electrostatic 
coupling strengths are below the threshold for charge renormalization.  

The resulting effective pair potentials have the same Yukawa form as predicted by 
linearized Poisson-Boltzmann theory and integral-equation theories.  Our expressions
are somewhat more general, however, by incorporating macroion density and excluded 
volume via the Debye screening constant.  As a quantitative test of accuracy, we have 
calculated structural properties from molecular dynamics simulations of the coarse-grained
model, taking the effective pair potentials as input.  Radial distribution functions
of binary mixtures of oppositely charged colloidal macroions are found to agree 
closely with corresponding results from Monte Carlo simulations of the primitive model.
For a highly asymmetric (colloid-nanoparticle) mixture, our results demonstrate that
nanoparticles can enhance electrostatic screening, thus weakening pair correlations,
in suspensions of charged colloids.
Assessing the range of validity of the theory will require further comparisons with 
primitive model simulations and experiments.

The one-body volume energy, which depends on the average density of the macroions,
can influence the phase behavior and other thermodynamic properties, especially in
deionized suspensions.
For binary colloidal mixtures, our analytical expression for the volume energy is 
similar to that derived from density-functional theory~\cite{Dijkstra2010}, 
but incorporates macroion excluded volume in a different manner.  
The volume energy also is an essential element required to extend to mixtures the 
charge renormalization theory previously developed for monodisperse colloidal 
suspensions~\cite{Denton2008,Lu-Denton2010}.

A subject for future work is the application of the effective interaction theory 
developed here to explore the structure and thermodynamic phase behavior of macroion
mixtures, including colloid-nanoparticle mixtures, distinguished by extreme 
size and charge asymmetries~\cite{ChungDentonUnpublished}.  Particularly interesting 
would be an investigation of the possibility of electrostatically driven bulk 
phase separation in deionized suspensions and a generalization to mixtures of a 
previously proposed charge renormalization theory~\cite{Denton2008,Lu-Denton2010}, 
which can significantly extend the range of validity of coarse-grained models 
to mixtures of highly charged macroions.

\begin{acknowledgments}
This work was supported by the National Science Foundation under Grant No. 
DMR-1106331.  Helpful discussions with Per Linse are gratefully acknowledged.
\end{acknowledgments}

\appendix
\section{Pressure calculation}\label{appPressure}
For our coarse-grained model of colloidal mixtures, the virial theorem for the pressure
must be generalized to account for the density-dependence of the effective 
pair potentials~\cite{Lu-Denton2010}.  To this end, the ensemble average in 
Eq.~(\ref{pMac}) can be written more explicitly as
\begin{equation}
\la \left(\pd{U}{V}\right)_{x_m,x_s} \ra =
-\la \frac{\cal V_{\rm{int}}}{3V}\ra + 
\la \left(\pd{U}{V}\right)_{x_m,x_s,\{\rv\}} \ra~,
\label{partialUpartialV}
\end{equation}
where the first term on the right-hand side involves the usual internal virial 
${\cal V_{\rm{int}}}$ and the partial derivative in the last term is taken for 
a fixed configuration of macroions $\{\rv\}$.  
For a mixture, the internal virial is
\begin{eqnarray}
{\cal V_{\rm{int}}} &=&
\sum_m \sum_{i<j}^{N_m} (1 + \kappa r_{ij})\veff{m}{m}(r_{ij})
\nonumber\\
&+& \sum_{m<n}\sum_{i=1}^{N_m}\sum_{j=1}^{N_n} 
(1 + \kappa r_{ij}) \veff{m}{n}(r_{ij})~.
\end{eqnarray}
Noting that $U$ depends implicitly on the volume through $\kappa$, we can write 
\begin{equation}
\left(\pd{U}{V}\right)_{x_m,x_s,\{\rv\}} 
= \left(\pd{U}{\kappa}\right)_{\{\rv\}} 
\left(\pd{\kappa}{V}\right)_{N_m,\Ns}~,
\end{equation}
where 
\begin{equation}
\left(\pd{\kappa}{V}\right)_{N_m,\Ns} = -\frac{\kappa}{2V(1-\eta)}
\end{equation}
and
\begin{eqnarray}
\left(\pd{U}{\kappa}\right)_{\{\rv\}} &=& \sum_{m} \sum_{i<j}^{N_m} 
f_m(r_{ij}) \veff{m}{m}(r_{ij}) 
\nonumber \\
&+&
\sum_{m<n}\sum_{i=1}^{N_m} \sum_{j=1}^{N_n}
f_{mn}(r_{ij}) \veff{m}{n}(r_{ij})
\end{eqnarray}
with
\begin{equation}
f_m(r_{ij}) = \frac{2 \kappa a_m^2}{1+\kappa a_m} - r_{ij}
\end{equation}
and 
\begin{equation}
f_{mn}(r_{ij}) = 
\frac{\kappa[a_m^{2} + a_n^{2}+\kappa (a_m+a_n)a_m a_n ]}
{(1+\kappa a_m)(1+ \kappa a_n)} - r_{ij}~.
\end{equation}

\bibliography{effBinary}

\end{document}